\documentclass[11pt]{amsart}
\usepackage[latin1]{inputenc}
\usepackage[cyr]{aeguill}
\usepackage{amsxtra}
\usepackage{amssymb}
\usepackage[all]{xy}
\newtheorem{theorem}{\sc Theorem}[section]
\newtheorem{proposition}{\sc Proposition}[section]

\newtheorem{corollary}{\sc Corollary}[section]
\theoremstyle{remark}
\newtheorem{remark}{\it Remark}[section]
\newtheorem{example}{\it Example}[section]
\newtheorem{definition}{\sc Definition}[section]
\font\tmsb=msbm10 at12pt
\font\smsb=msbm7
\font\ssmsb=msbm5
\newfam\msbfam
\textfont\msbfam=\tmsb
\scriptfont\msbfam=\smsb
\scriptscriptfont\msbfam=\ssmsb
\def \det{{\rm det\,}}
\def \Hom{{\rm Hom}}
\def \Ens{{\rm Ens}}
\def \Sp{{\rm Sp}}
\def \rank{{\rm rank\,}}
\def \RM{\mathbb {R}}%        corps des reels
\def \NM{\mathbb{N}}%        entiers naturels
\def \ZM{\mathbb{Z}}%        entiers relatifs
\def \CM{\mathbb{C}}%        nombres complexes
\def \HM {\mathbb{H}}

\def \d{\partial}%derivee partielle

\def\dt{\delta} 
\def\a{\alpha}
\def\b{\beta}
\def\e{\varepsilon}  

\def\p{\varphi}

\def\l{\lambda}

\def\s{\sigma}

\def \S{\Sigma}
\def \t{\tilde}

\def \to{\longrightarrow} 
\def \w{\wedge}
\def\2n{(\CM^{2n},0)}

\def \< {{\langle }}
\def \> {{\rangle }}

\newcommand{\Bt}{{\mathcal B}}

\newcommand{\Ft}{{\mathcal F}}

\newcommand{\Ht}{{\mathcal H}}

\newcommand{\Mt}{{\mathcal M}}
\newcommand{\Ot}{{\mathcal O}}

\newcommand{\Qt}{{\mathcal Q}}

\newcommand{\Cq}{{\CM_{\hbar}}}

\medskip
\begin{document}
\title[Analytic geometry and semi-classical analysis]{Analytic geometry and \\ semi-classical analysis}
\author[Mauricio D. Garay]{ Mauricio D. Garay}
%\author{\`A V.I. Arnold pour ses 70 ans et en remerciement de son enseignement}
\date{May 2007}
\address{SISSA/ISAS Via Beirut 2-4, 34014 Trieste, Italy}
\email{garay@sissa.it}
\thanks{\footnotesize 2000 {\it Mathematics Subject Classification:}
81Q15}
\keywords{Integrable systems,
Symplectic geometry, Lagrangian varieties,
Rayleigh-Schr\"odinger series.}
%%%%%%%%%%%%%%%%%%%%%%%%%%%%%%%%%%%%%%%%%%%%%%%%%%%%%%%%%%%%%%%%%
%%%%%%%%%%%%%%%%%%%%%%%%%%%%%%%%%%%%%%%%%%%%%%%%%%%%%%%%%
%%%%%%%%%%%%%%%%%% ABSTRACT%%%%%%%%%%%%%%%%%%%%%%%%%%%%%%
%%%%%%%%%%%%%%%%%%%%%%%%%%%%%%%%%%%%%%%%%%%%%%%%%%%%%%%%%
\begin{abstract}
We study deformation theory for quantum integrable systems and prove several
theorems concerning the Gevrey convergence and the unicity
of perturbative expansions.

\end{abstract}
\maketitle
\parindent=0cm
\vskip0.5cm
\begin{flushright} {\it {\`A} V.I. Arnold pour ses 70 ans.} \end{flushright}
\section*{Introduction}
The aim of this paper is to give an account on deformation theory for Lagrangian varieties and its applications to semi-classical
analysis. Semi-classical analysis, as initiated by Arnold and Maslov,
is the study of partial differential equations depending on a small parameter $\hbar$;
in its major part it was developed as a global study for real compact Lagrangian submanifolds.
Our aim is to explain the interplay between the complex topology of the complexified characteristic varieties and the spectral
theory of pseudo-differential operators, the bridge between
both approaches being given by deformation theory of Lagrangian varieties.\\
In standard spectral theory, the perturbative expansions for the spectrum are a purely formal object defined ad-hoc. In the more modern
approach of real $C^\infty$ semi-classical analysis such objects have non-formal meaning as
 perturbations of global objects but still a notion like the spectrum of the germ of an operator has usually no meaning.
The relationship between topology, analytic geometry and semi-classical analysis can only be established, if we make sense
of such an object.
So, our first goal will be to reformulate the standard spectral theory, so to attach to each germ of an operator
a spectrum which behaves naturally with respect to base changes, that is,
for which the perturbative expansion is the spectrum of an operator relatively to the base of a deformation.\\
For simplicity, we will start with the case of partial differential equations depending on one variable, that is, ordinary differential equations. In this case,
the correspondence between the topology of characteristic varieties, analytic geometry and semi-classical analysis
is more easily understood, essentially due to the fact that any plane curve is a Lagrangian submanifold.\\
Once the correspondence is established, we will essentially discuss deformations of Lagrangian varieties, problems of stability
and their relations to the topology of Lagrangian Milnor fibres.\\
The correspondence between analytic geometry and semi-classical analysis will constitute our first task,
it is summarised by the following table
\vskip0.5cm
 \begin{tabular}{||  p{6cm} |  p{6cm}||} \hline
 \begin{center} Analytic geometry \end{center}& \begin{center} Semi-classical analysis \end{center}\vskip0.2cm \\
 \hline
\vskip0.1cm
Coherence of the Lagrangian deformation module & \vskip0.1cm Convergence of perturbative expansions in the semi-classical limit \\ \hline
\vskip0.1cm
Freeness of the Lagrangian deformation module & \vskip0.1cm Uniqueness of the perturbative expansions in the semi-classical limit \\ \hline
\vskip0.1cm
Coherence of the quantised Lagrangian deformation module& \vskip0.1cm Gevrey-convergence of the perturbative expansions \\ \hline
\vskip0.1cm
Freeness of the quantised Lagrangian deformation module & \vskip0.1cm Uniqueness of perturbative expansions \\ \hline
\end{tabular}
\vskip0.5cm
For expository reasons, we concentrate ourselves in germs at a point. As most of our arguments are sheaf theoretic, the
neighbourhood of the point can be replaced by the neighbourhood of any compact subset with almost no modification (for instance neighbourhoods of periodic orbits in Hamiltonian systems). Also, to adapt the theorems to the real analytic
case is a rather trivial exercise. The reader interested only in the formal aspect of expansions and not on their analyticity
will easily recognise that, in that case, most statements hold with trivial proofs and without assuming any integrability or holonomicity condition.\\

V.I. Arnold's students all know his ability to transmit his passion for mathematics and for science in general.
For encouragements, criticism and teaching, I would like to express my gratitude.\\
Much of this work was elaborated at the university of Mainz,
I thank all members of the mathematical department and in particular D. van Straten for the rich and stimulating atmosphere created at the University and to C. Sevenheck for discussions on Lagrangian varieties.\\
Thanks also to Colin de Verdi\`ere who taught me semi-classical analysis and to B. Teissier who insisted that I write a text gathering scattered results, sometimes
published and sometimes simply exposed in seminars. 
\newpage
\enlargethispage{1cm}
\tableofcontents
\newpage
\section{The spectrum of an operator in the Heisenberg algebra}

%%%%%%%%%%%%%%%%%%%%%%%%%%%%%%%%%%%%%%%%%%%%%%%%%%%%%%%%%%%%%%%%%55
%%%%%%%%%%%%%%%%%%%%%%%%%%%%%%%%%%%%%%%%%%%%%%%%%%%%%%%%%%%%%%%%%%%%%%%%%%%%%%%%%%%%%%%%%%%%%%%%%%%%%%%%%%%%%%%%%%%%%%%%%
\subsection{The Rayleigh-Schr{\"o}dinger expansions}
Consider the unbounded operators $A,A^\dag$ on the Hilbert space $l^2(\NM)$ with canonical basis
$e_0=(1,0,0,\dots),$ $e_1=(0,1,0,0,\dots)$ defined
by
$$ A^\dag\, e_n=\sqrt{n+1}\, e_{n+1},\ A\, e_0=0,\ A \, e_n=\sqrt{n}\, e_{n-1},\ n>0.$$
The spectrum of the harmonic oscillator $H_0=\ A^\dag A$ consists of non-negative integer $\ZM_{\geq 0}$.
Heisenberg showed that for cubic and quartic polynomial perturbations $H_t=H_0+t P(A,A^\dag)$,
the eigenvalues admit formal asymptotic expansions $E_n(t)=\sum_k \a_n t^n$ with
$E_n(0)=n$ (\cite{Heisenberg}). These asymptotic expansions are now called 
{\em Rayleigh-Schr{\"o}dinger expansions}; they have no meaning in the standard spectral theory which axiomatises quantum mechanics.
When the degree of $P$ is at least $3$, the perturbation is therefore called a 
{\em singular perturbation} since it changes completely the qualitative behaviour of the spectrum.\\
Our aim is to interpret these expansions
as the spectrum of an operator relative to the base of a deformation and then prove the following theorem.
\begin{theorem}[\cite{quantique}]
\label{T::analytique}
{For any polynomial perturbation of a harmonic oscillator
$H_t=A^\dag A+t P(A^\dag,A)$, the associated Rayleigh-Schr{\"o}dinger expansions are asymptotic expansions of holomorphic
functions at the boundary of their holomorphy domain.}
\end{theorem}
This theorem will be useful to illustrate our approach and to motivate the concepts we shall introduce.
%%%%%%%%%%%%%%%%%%%%%%%%%%%%%%%%%%%%%%%%%%%%%%%%%%%%%%%%%%%%%%%%%%%%%%%%%%%%%%%%%%%%%%%%%%%%%%%%%%%%%%%%%%%%%%%%%%%%%%%%%%%%%%%%%%%%%%%%%%%%%%%%%%%%%
\subsection{Asymptotic expansions and Gevrey analyticity}
A sector in $\CM$ is an open subset of the type
$$S_{\theta,r}=\{ x \in \CM/ |Arg\, x|<\theta, |x|<r \}$$
with $\theta \in \RM_{>0}$.
An asymptotic expansion $\sum_{n \geq 0} a_n x^n$ is associated to a holomorphic function $f:S_{\theta,r} \to \CM$ if
$$\lim_{x \to 0}f^{(n)}(x)=n!a_n$$
The map 
$$B:\CM[[x]] \to \CM[[x]],\ \sum_{i \geq 0} a_{i}x^i \mapsto \sum_{i \geq 1} a_{i}\frac{x^{i-1}}{(i-1)!}.$$
  is called the  {\em Borel transform}. 
A formal power series is of  {\em Gevrey class} provided that its Borel transform is analytic.
\begin{theorem}[see \cite{Malgrange_resommabilite}]
Any formal power series of  Gevrey class is the asymptotic expansion associated to a holomorphic function.
\end{theorem}
Therefore in order to show that an asymptotic expansion is associated to a holomorphic function is is sufficient to show that it is of Gevrey class.
 %%%%%%%%%%%%%%%%%%%%%%%%%%%%%%%%%%%%%%%%%%%%%%%%%%%%%%%%%%%%%%%%%%%%

\subsection{The $\Qt$-algebra}
The (one-dimensional) universal Heisenberg algebra $\hat \Qt$  is the algebra of formal power series 
in $\CM$ generated by three elements $(a^\dag), a, \hbar$ and satisfying only one non-trivial commuting relation
$$[a,a^\dag]=\hbar.$$
Unlike the usual treatment, we do not consider the Weyl algebra where the commutation relation is $[A,A^\dag]=1$
with $A=a/\hbar$, $A^\dag=a^\dag$.\\
Following Pham, we define an analytic subalgebra $\Qt \subset \hat \Qt$ (\cite{Pham_resurgence}).\\
The total symbol of an element $f \in \hat \Qt$ is obtained by ordering the expansion of $f$ so that
$$ f=\sum_{i,j \geq 0}\a_{ij}(a^{\dag})^i a^j $$
and then by substituting the operators $a,a^\dag$ by commuting variables $x,y \in \CM$.\\
We define a Borel transform in $\hat \Qt$ by sending a map to its total symbol and then by taking the image of the total symbol
under the map
$$\CM[[\hbar,x,y]] \to \CM[[\hbar,x,y]],\ \sum_{i,j,k \geq 0} a_{ijk}x^iy^j\hbar^k
\mapsto \sum_{i,j,k \geq 0} a_{ij}x^iy^j\frac{\hbar^{k-1}}{(k-1)!}.$$
We define the vector subspace $\Qt \subset \hat \Qt$ to be the subspace
of operators which have a total symbol with an analytic Borel transform in a small neighbourhood of the origin:
$$f \in \Qt \iff B(f) \in \CM\{ x,y \}. $$
Here $\CM\{ x,y \}$ denotes the algebra of holomorphic function germs at the origin in $\CM^2$.\\
As remarked by Pham, that the vector space $\Qt$ is isomorphic to an algebra of analytic pseudo differential operators;
therefore it follows from results due to Boutet de Monvel-Kr{\'e}e that the vector subspace $\Qt \subset \hat \Qt$
is, in fact, a subalgebra \cite{Boutet_Kree,Pham_resurgence} (see also \cite{quantique}).\\
We denote by $\Cq \subset \Qt$ the operators in $\Qt$ which depend only on $\hbar$ (the constants).
In the above definitions, one might replace the field $\CM$ by any ring of formal power series.
In particular, the algebra $\Qt$ admits central extensions $\Qt\{ \l \}$, $\l=(\l_1,\dots,\l_k)$, obtained by replacing the ring $\CM$
with $\CM[[\l ]]$. We denote by $\Cq \{ \l \}$ analytic series
with coefficient in $\Cq$, i.e., formal power series of the type 
$$ \sum_{i \in \ZM_{\geq 0},j \geq 0}\a_{ij}\l^i \hbar^j $$
having an analytic Borel transform in $\hbar$. When $k=1$, we shall often write $t$ or $z$ instead of $\l$.
 %%%%%%%%%%%%%%%%%%%%%%%%%%%%%%%%%%%%%%%%%%%%%%%%%%%
\subsection{Functoriality requirements}
Any element $H \in \Qt\{ \l \}$ defines a $\Cq\{ z,\l \}$-module structure on the algebra $\Qt\{ \l \}$ by putting 
$\psi \circ H=\sum_{n \geq 0} \a_n H^n$ with $\psi(z)=\sum_{n \geq 0} \a_n z^n$, $\a_n \in \Cq\{ \l \}$.\\
In order to relate Theorem \ref{T::analytique} to deformation theory in the Heisenberg algebra,
we need to construct a spectral theory functorial with respect to inverse images and automorphisms, i.e.,
we want to construct a mapping $Sp:\Qt\{ \l \} \to \Ens(\Cq\{ \l \})$ such that:
 \begin{itemize}
\item $Sp(\p(H))=Sp(H)$ for any automorphism $\p \in Aut(\Qt \{ \l \})$ with
$\p(\l=0,\cdot)=Id$,
\item $Sp(\psi \circ H)=\psi (Sp(H))$, $\psi \in \Cq\{ \l,z \}$. 
\end{itemize}
Here $\Ens$ denotes the category of subsets and $Id$ the identity mapping.\\
To construct the spectrum, we first need to construct the space on which the operators act.
%%%%%%%%%%%%%%%%%%%%%%%%%%%%%%%%%%%
\subsection{Bargmann's representation for the harmonic oscillator}
Consider the pre-Hilbert space $\Bt$ of holomorphic function in $\CM$ which are square-integrable for the
Gaussian measure:
$$\int_{\CM}| f(z)|^2 e^{-z\bar z} dz \w d\bar z <+\infty. $$
The Weyl algebra of linear partial differential operators act on the vector space $\Bt$, its generators $A=\d_z$
and $A^\dag=z$ satisfy the commutation relation $[A,A^\dag]=1$. The spectrum of the operator
$A^\dag A$ is the set $\ZM_{\geq 0}$ and the $z^n$'s form a generating system of eigenvectors.\\
Our aim is to do a similar local analytic construction involving the semi-classical parameters $\hbar$ and allowing the
operators to be given by infinite analytic series and not just polynomials.
%%%%%%%%%%%%%%%%%%%%%%%%%%%%%%%%%%%%%%%%%%%%%%%%%%%%%%%%%%%5
\subsection{The Hilbert module $\Ht$}
The quotient $\Ht=\Qt/\Qt a$ of  $\Qt$ by the left-ideal generated by $a$ is isomorphic to the space
$\Cq \{ z \}$. Indeed, modulo the ideal $\Qt a$ any operator can be written as a series $\sum_{n}\a_n (a^\dag)^n$ and the isomorphism
is given by the map
$$ \Ht \to \Cq\{ z \},\, \sum_{n}\a_n (a^\dag)^n \mapsto \sum \a_n z^n.$$
Consider the canonical projection $\pi:\Qt \to \Ht$.
The left multiplication on $\Qt$ by an element $H \in \Qt$ gives a commutative diagram
$$\xymatrix{\Qt \ar[r]^{\cdot H} \ar[d]^{\pi} & \Qt \ar[d]^\pi \\
                 \Ht \ar[r]^{\rho(H)} & \Ht }
$$
Thus, the left multiplication in $\Qt$ induce a representation
$$\rho:\Qt \to \Hom_{\Cq}(\Ht,\Ht)$$
which can easily be shown to be injective.
Therefore, we can identify the algebra $\Qt$  with a subalgebra acting on $\Ht$ (\cite{quantique}).\\
Eigenvectors $v \in \Ht$ of $H$ (resp.  {\em eigenvalues} $E \in \Cq$) are defined by the condition
$$ \rho(H)v=Ev. $$
Here and in the sequel, we shall simply write $ Hv$ instead of $\rho(H)v $.
Using the $\Cq$-module $\Ht \approx \Cq\{ z \}$ isomorphism, the representations $\rho$ is defined by
$$\rho(a):\Cq\{ z \} \to \Cq\{ z \},\ \psi \mapsto \hbar \d_z \psi$$
and
$$\rho(a^\dag):\Cq\{ z \} \to \Cq\{ z \},\ \psi \mapsto z \psi.$$
Similar consideration holds for the formal Heisenberg algebra.
More generally, if $l:\CM^2 \to \a x+\b y$ is a non-vanishing linear form, the associated Hilbert module
is $\Ht_l=\Qt/\Qt(\a a^\dag+\b a)$, the choice of such a linear form is called a choice of {\em polarisation}. 
\begin{example}
Consider the harmonic oscillator $H=a^\dag a$.
We have $H (a^\dag)^n=n\hbar (a^\dag)^n \ ({\rm mod} \ \Qt a)$ therefore the
projection of the $(a^\dag)^n$'s in $\Ht$ form a basis of eigenvectors for $H$. Via the above correspondence,
we may identify the operator $H$ with
$\hbar z \d_z$ acting on $\Cq \{ z \}$, the above-mentioned basis of eigenvectors correspond to the function germs
$z^n$.
\end{example}
%%%%%%%%%%%%%%%%%%%%%%%%%%%%%
\subsection{Relative spectrum and perturbative expansion}
We now consider the case with parameters.\\
In the space $\Ht\{ \l \}:=\Qt\{ \l \}/\Qt\{ \l \} a$ we cannot take the above definition for the eigenvectors, since already in example of the harmonic oscillator, perturbative expansions of eigenvectors are, in general, neither holomorphic nor meromorphic. Therefore, we shall say that $\psi \in \Ht\{ \l \}$ is an eigenvector of $H \in \Qt\{ \l \}$ if there exists
$E \in \Cq \{ \l \}$ such that
$$H(\hbar t,\cdot) \psi(\hbar t,\cdot)=E(t,\cdot) \psi(\hbar t,\cdot). $$ 
Consequently, if $\psi(t,\cdot)$ is an eigenvector then  $\psi(t/\hbar,\cdot)$ is the corresponding perturbative expansion.
Similar consideration apply in the formal case.
Heisenberg result's on the existence of perturbative expansion can be stated as follows.
\begin{theorem}[\cite{Heisenberg}] For any perturbed anharmonic oscillator $H \in \Qt\{ t \}$, the restriction to $t=0$
induces a bijection 
$$\widehat \Sp(H) \to \widehat \Sp(H_0),\ H_0=H(t=0,\cdot) \in \Qt $$
\end{theorem}
%%%%%%%%%%%%%%%%%%%%%%%%%%%%%%%%%%%%%%%%%%%
\subsection{The analytic spectrum is functorial}
By {\em Heisenberg equations}, we mean a non-autonomous
evolution equation of the type
$\dot F=\frac{i}{\hbar}[ F,H], \ F, H \in \Qt\{ t \}$
where the dot denotes the derivative with respect to $t$.
The functoriality of the spectrum is based on the following result.
\begin{proposition}[\cite{quantique}]
\label{P::integration}
{For any $H \in \Qt\{ t \}$, there exists a unique operator 
$U \in \Qt \{ t \}$ satisfying the equation $\dot U=HU$ with initial condition $U(t=0,\cdot)=1$.
The  automorphism $\p \in Aut(\Qt \{ t \})$
$$\p:\Qt \{ t \} \to \Qt\{ t \}, f \mapsto
U(\frac{t}{i\hbar})f U^{-1}(\frac{t}{i\hbar}),\ U \in \Qt\{ t \},$$
integrates the Heisenberg equations of $H \in
\Qt\{ t \}$, that is:
$$\frac{d}{dt}\p(f) =\frac{i}{\hbar}[\p(f),  H ],\ \forall f \in \Qt.$$}
\end{proposition}
\begin{corollary}[\cite{quantique}]
\label{C::invariant}
For any automorphism $\p $ of $\Qt \{ t \}$, any mapping $\psi \in \Cq\{t,z \}$, and any
germ $H \in \Qt\{t \}$, we have
$$\Sp(\p(H))=\Sp(H),\ \Sp(\psi(H))=\psi(\Sp(H))$$
and similarly for the formal spectrum.
\end{corollary}

%%%%%%%%%%%%%%%%%%%%%%%%%%%%%%%%%%%%%%%%%%%%%%
%%%%%%%%%%%%%%%%%%%%%%%%%%%%%%%%%%%55
\subsection{Digression: the Trace of an operator in $\Qt$}
First, we construct a non-degenerate $\Cq$-sesquilinear product in $ \Ht$.\\
The hermitian conjugate
$$\Qt \to \Qt,\ H \mapsto H^\dag$$
is the unique  $\Cq$-antilinear  mapping for which the conjugate of $(a^\dag)^ia^j$ is $(a^\dag)^j a^i$.\\
Let us define the map
$$ev:\Qt \times \Qt \to \Cq,\ (H,G) \mapsto s(H^\dag G)_{x=y=0}$$
where $s$ denotes the total symbol.\\
The hermitian product $\< \cdot,\cdot \>  $ on $\Ht$ is defined by the commutative diagram
$$\xymatrix{ \Qt \times \Qt \ar^{ev}[r] \ar^\pi[d]& \Cq \\
               \Ht \times \Ht \ar_{\< \cdot, \cdot \> }[ru]}
$$

The space $\Ht$ satisfies most axioms of Hilbert spaces except that it is not a $\CM$-vector space but a $\Cq$-module. This drawback turns out to be an important
advantage for considering traces of operators.\\
Let us denote by  $| n \>  \in \Ht$, the class of the operator $(a^\dag)^n$
and by $ H \mid n \> $ its image under an operator $ H $.

\begin{theorem}[\cite{quantique}]
The trace mapping $Tr:\Qt \to \Cq,\ H \mapsto \sum \< n | H | n \> $
is defined for all operators.
\end{theorem}
This contrasts of course with the standard theory where so few operators have a trace.
To illustrate the construction, let us consider the case of an harmonic oscillator $H=a^\dag a$.\\
We get
$$Tr(H)=\sum_{n >0}\< n | H | n \> =\sum_{n>0}n\hbar \< n | n \> .$$
Using the equality $\< j | j \> =j!\hbar^j$, we deduce that
$Tr(H)=\sum_{n > 0}n\, n!\hbar^{n+1}$.
the Borel transform of the trace is therefore given by 
$$B Tr(H)=\sum_{n > 0}n\hbar^n=\frac{\hbar}{(1-\hbar)^2}.$$
For $ \hbar=1 $, the function has a singularity corresponding to the fact that the operator is not of trace class in the standard theory:
$$ \lim_{\hbar \to 1}\hbar\sum_{n > 0}n\hbar^n= \lim_{\hbar \to 1}\frac{\hbar}{(1-\hbar)^2}=+\infty .$$
The trace allows us to define a one-dimensional quantum field theory for the scalar field using the standard Hamiltonian formalism (\cite{Zinn}). The trace is invariant under automorphisms which preserve the polarisation, it does not satisfy $Tr(AB)=Tr(BA)$, this is in accordance with the fact that the correlators do not in general correspond to physical quantities.
%%%%%%%%%%%%%%%%%%%%%%%%%%%%%%%%%%%%%%%%%%
\section{The quantum Morse lemma}
%%%%%%%%%%%%%%%%%%%%%%%%%%%%%%%%%5
\subsection{Statement of the theorem}
The complex Morse lemma in two variables states that
any holomorphic function germ of the type
$f:(\CM^2,0) \to (\CM,0), (x,y) \mapsto x^2+y^2+o(|x|^2+|y|^2)$ can be taken back to its quadratic part
$(x,y)\mapsto x^2+y^2$ by a biholomorphic change of coordinates. The following result is an analogous result in $\Qt$
\begin{theorem}[\cite{quantique}]
\label{T::quantique}
Let $H_0 \in \Qt$ be an operator which total symbol has a non-degenerate quadratic part.
For any deformation $H \in \Qt \{ t \}$ of $H_0$, i.e. $H( t=0,\cdot)=H_0$, there exist an automorphism
$\p \in Aut(\Qt\{ t \})$ and a map $\psi \in \Cq \{ t,z \}$ such that
$$\p(H)=\psi \circ H_0.$$
\end{theorem}
The Gevrey convergence of the Rayleigh-Schr\"odinger series is a consequence of this theorem and of the functoriality
property of the spectrum  (Corollary \ref{C::invariant}).\\
Using the path method, we will see that the above theorem has the following corollary.
\begin{corollary}[\cite{quantique}]
\label{C::quantique}
For any operator $H_0 \in \Qt$ which total symbol has a non-degenerate quadratic part there exist an automorphism
$\p \in Aut(\Qt)$ and a map $\psi \in \Cq \{ z \}$ such that
$$\p(H_0)=\psi \circ (a^\dag a).$$
\end{corollary}
Corollary \ref{C::quantique} was obtained by Helffer and Sj{\"o}strand with the additional assumption that the operator is self-adjoint 
(\cite{Helffer_Sjostrand}, Th\'eor\`eme b1 and Th\'eor\`eme b6).
To explain Theorem \ref{T::quantique} and its proof, we first consider its semi-classical limit.
%%%%%%%%%%%%%%%%%%%%%%%%%%%%%%%%%%%%%%%%%%%%%%%%%%%%%%%%%%%%%%%%%%%%%%%%%%
\subsection{The isochore Morse lemma}
The principal symbol
$$\s:\Qt \to \CM\{ x,y \},\ \CM\{ x,y \} \approx \Qt/\hbar \Qt$$ maps the operator
$\sum_{jk}(\a_{jk}+o(\hbar))(a^\dag)^ja^k$ to the analytic expansion
$\sum_{jk}\a_{jk}x^jy^k$ with $\a_{jk} \in \CM$.\\
The Poisson bracket on $\CM^2 $ associated to the symplectic form $dx \w dy$ satisfies the equality
$$ \{ \s(f),\s(g) \}=\s(\frac{1}{\hbar}[f,g])$$
Therefore, any automorphism $\hat \p \in \Qt$ defines the germ of a symplectomorphism $\p$ in $\CM^2$ by
$$(\CM^2,0) \to (\CM^2,0),\ (x,y) \mapsto (\s(\hat \p(a)),\s(\hat \p(a^\dag)). $$
The {\em unfolding} associated to a deformation $F:(\CM^k \times \CM^2,0) \to (\CM,0)$ of a holomorphic function germ
$f:(\CM^2,0) \to (\CM,0)$ is the map 
$$\t F:(\CM^k \times \CM^2,0) \to (\CM^k \times \CM,0),\ (\l,x,y) \mapsto (\l,F(\l,x,y)).$$
Using the above process of semi-classical approximation to Theorem \ref{T::quantique}, we get the following result.
\begin{theorem}
\label{T::Morse_isochore}
Any deformation $F:(\CM \times \CM^2,0) \to (\CM,0)$
of a holomorphic function germ $f:(\CM^2,0) \to (\CM,0)$ having a non degenerate quadratic part is isochore trivial, i.e., there exists holomorphic
map germs $\p:(\CM \times \CM^2,0) \to (\CM \times \CM^2,0)$, $\psi:(\CM \times \CM,0) \to
(\CM,0)$ such that the following diagram commutes

$$\xymatrix{(\CM \times \CM^2,0) \ar[r]^{\t F} \ar[d]^\p &
(\CM \times  \CM,0) \ar[d]^\psi \\
(\CM^2,0) \ar[r]^f  & (\CM,0) }
$$ 
and $\p^* dx \w dy=dx \w dy$ where $(x,y)$ are coordinates in $\CM^2$. Here $\t F$ denotes the unfolding of $F$.
\end{theorem}
This theorem has the following corollary, which is due to Vey.
\begin{theorem}[\cite{Vey_isochore}]

\label{T::Vey}Any holomorphic map-germ  $f:(\CM^2,0) \to (\CM,0)$
having a non-degenerate quadratic part can be written as
$$f \circ \p (x,y)=\psi(xy).$$
where   $\p:(\CM^2,0) \to (\CM^2,0)$ is a symplectomorphism, i.e., $\p^*dx \w dy=dx \w dy$ and $ \psi:(\CM,0) \to (\CM,0)$ is biholomorphic.
\end{theorem}
In the statement of the theorem the quadratic form $xy$ can be replaced by any non degenerate quadratic form.
\begin{remark}
Conversely, using the isochore Morse lemma, Colin de Verdi{\`e}re-Parisse and Sj{\"o}strand proved
a quantum Morse lemma for formal power series in $\hbar$
(\cite{Parisse,Sjostrand}). This weaker statement does not give any information
on the Gevrey analyticity of perturbative expansions.
\end{remark}

Both results of this subsection (and also these of the preceding one) are related by the {\em path method}, namely
Theorem \ref{T::Morse_isochore} implies Theorem \ref{T::Vey}. Indeed, let us define the one parameter
family of holomorphic function germs
$f_t=(1-t)f+t d^2 f$ depending on the parameter $t \in [0,1]$.
According to Theorem \ref{T::Morse_isochore}, for any $t_0 \in [0,1]$, the germ of $f_t$ at
$t=t_0, x=0$ is a trivial deformation of $f_{t_0}$. Therefore,
$f_{t_0+\e}$ is conjugated to $f_{t_0}$ by a symplectomorphism germ in $(\CM^2,0)$ and a biholomorphic map in $(\CM,0)$,
for any $\e$ small enough. This shows that $f_0$ and $f_1$ are conjugated and therefore Theorem \ref{T::Vey} follows from Theorem \ref{T::Morse_isochore}.\\

To conclude the subsection, let us point out that there exists a real
$C^\infty$ isochore Morse lemma due to Colin de Verdi{\`e}re and Vey (\cite{Vey_Colin}). One can then introduce a micro-local spectrum on a compactification
of the curve $xy=0$, the local study is then replaced by a real compact global one (\cite{Parisse,Sjostrand}); this is the now standard approach in real $C^\infty$ semi-classical analysis. 
%%%%%%%%%%%%%%%%%%%%%%%%%%%%%%%%%%%

\section{A small review on local Gauss-Manin theory}
The results of this section are valid for hypersurface singularities and more generally for isolated complete intersection
singularities; in view of our applications, we will restrict ourselves to the case of plane curves. 
The relation of Gauss-Manin theory with Rayleigh-Schr\"odinger expansions will be explained in the next section.
%%%%%%%%%%%%%%%%%%%%%%%%%%%%%%%%%%%%%%%%%%%%%%%%%%%%%%%%%%%%%%%%%%%%%%%%%%%%%%%%%%%%5
\subsection{Topology of plane curves singularities}
A {\em standard representative}\footnote{Sometimes called a Milnor representative or a good representative.}  $ f:X \to S$ of a holomorphic map germ $f:(\CM^2,0) \to (\CM,0)$ with an isolated critical point is
a holomorphic representative constructed as follows.\\
Let $g:Y \to T$ be a Stein representative of $f$ such that the spheres $S_{\e}$ centred at the origin of radius $\e \leq \e_0$ are transverse to the zero fibre of $g$.\\
By transversality, we may chose a closed neighbourhood of the origin $S \subset T$ such that the fibres of $g$ above $S$ are transverse to the sphere $S_{\e_0}$;
we put $X=g^{-1}(S) \cap B_{\e_0}$ where $B_{\e_0}$ denotes the ball centred at the origin of radius $\e_0$.
Remark that we abusively denote the standard representative and the germ by the same letter.
\begin{theorem}[\cite{Mil}]
The restriction of the map $ f:X \to S$ above the complement of the origin is a locally trivial $C^\infty$ fibre bundle, which does not depend, up to isomorphism, on the choice of $ f$.
 \end{theorem}

The fibration associated to a holomorphic function germ is called the {\em Milnor fibration}, the fibre of which is an open Riemann surface. Any open Riemann surface as the homotopy type of a bouquet
of circles, the following theorem indicates the number of such circles. 
\begin{theorem}[\cite{Palamodov,Mil}] The Milnor fibre associated to a holomorphic map germ  $f:(\CM^2,0) \to (\CM,0)$ has the homotopy
type of a bouquet of $\mu(f)$-circles with $\mu(f)=\dim\, \Ot_{\CM^2,0}/(\d_x f,\d_y f)$
\end{theorem}

The algebra $\Ot_{\CM^2,0}/(\d_x f,\d_y f)$ is called the {\em Milnor algebra}, the ideal $(\d_x f,\d_y f) \subset \Ot_{\CM^2,0}$ is called the {\em Jacobian ideal}
of $f$. When the context makes it clear, we shall simply denote by $\mu$ the Milnor number of a given function.

\begin{example}
For the holomorphic function germ $f:(\CM^2,0) \to (\CM,0)$ defined by $f(x,y)=x^2+y^2$, the Milnor algebra is generated by the class of $1$. The Milnor fibre taken above a real positive value 
retracts on its real part and has the homotopy type of a circle.
\end{example}

%%%%%%%%%%%%%%%%%%%%%%%%%%%%%%%%%%%%%%%%%%%%%%%%

\subsection{The relative de Rham complex}
We keep the notations of the previous subsection.
To a holomorphic map-germ $f$, we associate the {\em relative de Rham complex}  $\Omega^\cdot_{X/S}$ (\cite{Grothendieck_deRham}).
This complex is defined by
$\Omega^k_{X/S}=\Omega^k_X/df \w \Omega^{k-1}_X$ and the differential is induced by the exterior derivative of the de Rham complex $\Omega^\cdot_X$.\\
The direct images $\RM^k f_*\Omega^\cdot_{X/S}, k=0,1,$ of this complex are defined by the presheaves
$$U \mapsto \HM^k(f^{-1}(U),\Omega^\cdot_{X/S}),\ k=0,1$$
As $X$ is contained in a Stein subset, this hypercohomology spaces are obtained by restricting global sections to $U$:
$$ \HM^k(f^{-1}(U),\Omega^\cdot_{X/S}) \approx H^k(\Omega^\cdot_{X/S}(f^{-1}(U))) \approx H^k(\Omega^\cdot_{X/S}(X))_{|f^{-1}(U)}.$$
For instance, $\RM^0 f_*\Omega^\cdot_{X/S}=\Ot_S$, indeed if $dg=adf$ for some $a  \in \Ot_X(U)$
then $g \in \Ot_X(U)$ is constant along the fibres of $f$ and therefore can be written as a holomorphic function of $f$.\\
When $U$ does not contain the origin, the $\Ot_S(U)$-module $\RM^1 f_*\Omega^\cdot_{X/S}(U)$
can be identified with the space of holomorphic sections above $U \subset S$ of the fibration
$$\bigcup_s H^1(f^{-1}(s),\CM) \to S \setminus \{ 0 \}.$$
This is due to the fact that
\begin{enumerate}
\item the restriction of a differential form of the type $df \w \cdot$ to a fibre $\{ f={\rm constant} \}$ vanishes,
\item there is a relative Poincar\'e Lemma stating that the
relative de Rham complex is a resolution of the sheaf $f^{-1}\Ot_S$.
\end{enumerate}
This observation is due to Grothendieck (\cite{Grothendieck_deRham}).\\
If $U$ contains the origin there is no such a straightforward statement; nevertheless by the general philosophy of singularity theory, for $s \neq 0$, we expect a relation 
between the cohomology spaces
$$(\RM^1f_*\Omega^\cdot_{X/S})_s\approx H^1(f^{-1}(s),\CM) \otimes \Ot_{S,s} \approx \Ot_{S,s}^\mu $$
 which is a global object on a fibre and the cohomology space $H^1(\Omega^\cdot_{X/S,0})$ which is a local object defined for the germ of $f$ at the origin. This is the content of the following theorem due to Brieskorn and Deligne.
 \begin{theorem}[\cite{Br3}]
\label{T::Brieskorn_Deligne}
The cohomology space  $H_f=H^1(\Omega^\cdot_{X/S,0})$
associated to a function germ with an isolated critical point
$f:(\CM^2,0) \to (\CM,0)$ is a free $\CM\{ f \}$-module of rank $\mu$,
where  $\mu$ denotes the Milnor number of $f$. Moreover, there is a canonical isomorphism $(\RM^1f_*\Omega^\cdot_{X/S})_0 \approx H_f$.
\end{theorem}
The theorem holds true for isolated hypersurface singularities, in that case the freeness of the module $H_f$, conjectured by Brieskorn, was proved by Sebastiani
(\cite{Sebastiani}). 
\begin{example}
 Consider the function germ $f:(\CM^2,0) \to (\CM,0)$ defined by the formula   $f(x,y)=x^2+y^2$.
The class of the differential $\a=fydx$ is a coboundary of the relative de Rham complex, indeed: 
$$d(fydx)=df \w ydx +f dy \w dx $$
and $2fdy \w dx=df \w (ydx-xdy) \in df \w \Omega^1_X(X)$.\\
The Milnor number of $f$ equals one. The one form $ydx$ is not a coboundary thus the class of the differential
$\a=fydx$ is not divisible by $f$. Theorem \ref{T::Brieskorn_Deligne} shows that the module $H_f$ is generated by the class of $\a$.
\end{example}

%%%%%%%%%%%%%%%%%%%%%%%%%%%%%%%%%%%%%%%%%%%%%%%%%%%%55

%%%%%%%%%%%%%%%%%%%%%%%%%%%%%%%%%%%%%%%%%%%%%%%%%%%%%%%%%%%%%%%%%%%%%%%{\`u}

\subsection{The Brieskorn lattice}
Another way to construct holomorphic section of the cohomological Milnor bundle $\bigcup_s H^1(f^{-1}(s),\CM) \to S \setminus \{ 0 \}$
is to take the Gelfand-Leray residue of a two form on $X$.
We consider only differential with simple poles along the graph of $f$, the more general case is not difficult and leads
to Pham's construction of the micro-local Gauss-Manin module (\cite{Pham_GM}).

\begin{example} Consider a standard representative $f:X \to S$
of the germ  $f(x,y)=x^2+y^2$. To the holomorphic 2-form $\omega=dx \w dy$, we associate
the cohomology classes 
$$\omega_s=[\frac{dx \w dy}{d f}] \in H^1( f^{-1}(s),\CM) .$$
\end{example}
The holomorphic 2-form $\omega'=\omega+df \w d g$ induces the same section as $\omega$ of the cohomological Milnor bundle
for any holomorphic function $g:X \to S$. We define the Brieskorn lattice of $f$ by
$$H''_f=\Omega^2_X/df \w d \Ot_{\CM^2,0}.$$

\begin{theorem}[\cite{Br3}] For any holomorphic function
$f:(\CM^2,0) \to (\CM,0)$ with an isolated critical point at the origin, the Brieskorn lattice
is a free module of rank $\mu(f)$ and the cokernel of the map
$$H_f \to H''_f,\, \a \mapsto \a \w df$$
is a $\mu$-dimensional vector space.
\end{theorem}
\begin{example} Let us come back to the previous example. The holomorphic 2-form $dx \w dy \in \Omega^2(X)$ induces a class in $H''_f$ which is non zero and not divisible by $f$, therefore
it generates the Brieskorn lattice.
In particular, the restriction of the form $dx \w dy/df$ to any Milnor fibre generates its de Rham cohomology
space.
\end{example}
%%%%%%%%%%%%%%%%%%%%%%%%%%%%%%%%%%%%%%%%%%%%%%%%%%%%%%555
\subsection{Extension to deformations of plane curve singularities}
The notions that we reviewed for plane curves singularities,
extend naturally to deformations of plane curves singularities, and
in fact to any isolated complete intersection singularity. The corresponding finiteness and freeness results were proved
by Greuel (\cite{Greuel}).\\
Associated to the unfolding $\t F$ of $F$ there is a relative de Rham complex and a Brieskorn lattice
$$H''_F=\Omega^{k+2}_X/dF \w d\l \w d \Ot_{\CM^{k+2},0}$$
where we used the notation $d\l=d\l_1 \w d\l_2 \w \dots d\l_k$.
\begin{theorem}[\cite{Br3,Greuel}]
\label{T::Brieskorn}
 For any deformation
$F:(\CM^k \times \CM^2,0) \to (\CM,0)$ of a function-germ $f:(\CM^2,0) \to (\CM,0)$
with an isolated critical point at the origin, the Brieskorn lattice
is a free $\Ot_{\CM^{k+1},0}$-module of rank $\mu(f)$.
\end{theorem}
\begin{example}
\label{E::Brieskorn}
Consider a deformation $F:(\CM^3,0) \to (\CM^2,0),\ (t,x,y) \mapsto x^2+y^2+t g(x,y)$ where $g$ is an arbitrary
holomorphic function germ and take a standard representative
$\t F:X \to S, (t,x,y) \mapsto (t,F(t,x,y))$ of its unfolding.
The holomorphic 3-form $dx \w dy \w dt \in \Omega^3(X)$ generates the Brieskorn lattice:
$$H''_F=\CM\{ t,F \} [dx \w dy \w dt] $$
In particular, the restriction of the 1-form $dx \w dy \w dt /(dF \w dt)$ to any Milnor fibre of the unfolding
$\t F$ generates its first de Rham cohomology space.
\end{example}
%%%%%%%%%%%%%%%%%%%%%%%%%%%%%%%%%%%%%%%%%%%%%%%%%%%%%%%%%%%%%%%

\section{Proof of the theorems in the semi-classical limit}
\label{S::proof}
We have shown that certain analyticity properties of perturbative expansions follow from
deformation theory for operators. We now show that these theorems can be obtained as by-products of finiteness
results for the corresponding deformation module.
In the semi-classical limit, the deformation module is the Brieskorn lattice associated to the symbol of the deformation
as we shall now prove.

%%%%%%%%%%%%%%%%%%%%%%%%%%%%%%%%%%%%%%%%%%%%%%%%%%%%%%%%%%%%%%%%%

\subsection{The Moser-Poincar{\'e}-Thom path method}
In order to prove Morse Lemma-type theorems, we use the {\em path method}. This method consist in connecting our two objects by a path (for us the operator and its
quadratic part) and then prove that at each time the deformation is trivial. This is the method we employed in the second formulation of the isochore Morse lemma (Theorem \ref{T::Morse_isochore}).
Then, we replace the exact condition by an infinitesimal one: to solve the equation
 $$\psi_t \circ f_t \circ \p_t=f ;$$
we differentiate with respect to $t$ and multiply the result by $\p_t^{-1}$ on the right and by $\psi_t^{-1}$ on the left; we get an equation of the type 
$$a_t(f_t)+L_{v_t}f_t=g$$
where $g$ is a given holomorphic function germ and $v_t,a_t$ are then unknown.\\
If this equation is solvable, then the maps   $\psi_t,\p_t$ are obtained by integrating $v_t$ and $a_t$ along the parameter $t$.
This is the method used by Thom for proving the Morse lemma (\cite{AVG}).
 %%%%%%%%%%%%%%%%%%%%%%%%%%%%%%%%%%%%%%%%%%%%%%%%%%%%%%%%%%%%%%%{\`u}
\subsection{Peculiarity of the isochore problem}
The one-parameter family of maps $\p_t$ preserves the 2-form $dx \w dy$, therefore the non-autonomous vector field $v_t$
is Hamiltonian. Indeed, by differentiating the equality $\p_t^* dx \w dy=dx \w dy$ with respect to $t$, we get
 $L_{v_t} dx \w dy=0$, which by Cartan formula  $L_X=i_X d+d i_X$ implies that the one-form $i_{v_t} dx \w dy$ is closed; finally by
the Poincar\'e-de Rham lemma,
we get that $i_{v_t} dx \w dy=d H_t$ for some holomorphic function germ $H_t \in \CM \{ t,x,y \}$.\\
This shows that the infinitesimal condition reads
$$a_t(f_t)+\{ H_t,f_t \}=g.$$
Multiplying both members by the 3-form $\omega=dx \w dy \w dt$, we get the equation
$$a_t(f_t)\omega+d H_t \w d f_t \w dt=\a,\ \a=g \omega.$$
To get an idea of what this equation means, let us divide both members by $dt$ and take $t=0$, we get
$$a(f)\omega_0+d H_0 \w d f =\a_0,\ \omega_0=dx \w dy$$
which can be written as
$$a(f)[\omega_0]=[\a_0].$$
were $[ \cdot ]$ denotes the class in the Brieskorn lattice of $f$.\\
In other words the Brieskorn lattice can be identified with the space of infinitesimal deformations of $f$
modulo the one given by infinitesimal symplectic change of coordinates.\\
Now, Theorem \ref{T::Brieskorn}
implies that the Brieskorn lattice of $f_t$ is a free $\CM\{ t,f_t \}$-module
of rank one and  it is therefore generated by the class of $\omega$ (see Example \ref{E::Brieskorn}).\\
Thus, the infinitesimal equation can be solved, this shows
that the Vey isochore theorem (Theorem \ref{T::Vey}) is a consequence of Theorem \ref{T::Brieskorn}.\\
Remark that only the finiteness property is sufficient to guarantee the convergence of the expansions.
The freeness of the Brieskorn lattice (with parameters) implies that the function $a_t$ is unique and therefore so is the map $\psi_t$. In this way, we have a dictionary
 \begin{enumerate}
\item coherence of the Brieskorn lattice $\iff$ convergence of perturbative expansions in the semi-classical limit,
\item freeness of the Brieskorn lattice $\iff$ unicity of perturbative expansions in the semi-classical limit,
\end{enumerate}
as announced in the introduction (taking into account the fact that here the Lagrangian deformation module is the Brieskorn lattice
with parameters).
%%%%%%%%%%%%%%%%%%%%%%%%%%%%%%%%%%%%%%%%%%%%%%%%%%%%%%%%%%%%%%%%%%%{\`u}
\subsection{The versality theorem}
A deformation $F :(\CM^k \times \CM^2,0) \to ( \CM,0) $ of a holomorphic function germ $f:(\CM^2,0) \to (\CM,0)$
is called {\em RL-versal} if for any other deformation $G:(\CM^j \times \CM^2,0) \to (\CM,0)$ of $f$, there exists a commutative diagram
of holomorphic maps
\begin{equation}
\label{E::commutative}
\xymatrix{(\CM^j,0)\ar^-{i_j}[r] \ar[d]& (\CM^j \times \CM^2,0)\ar[r]^-{\t G} \ar[d]^-\p & (\CM^j \times \CM,0) \ar[d]^-{\psi}\\
          (\CM^k,0)  \ar^-{i_k}[r]  &   (\CM^k \times \CM^2,0)  \ar[r]^-{\t F} & (\CM^k \times \CM,0) \\ }
\end{equation}
 between the unfolding $\t F,\t G$ of the deformations. Here $i_j$ and $i_k$ denote the inclusions.\\
If the map $\p$ can be chosen so to preserve the two-form $dx \w dy$, the deformation $F$ is called {\em isochore versal};
it is called {\em isochore miniversal} in case the number of parameter on which depend the versal deformation is minimal.
 The following theorems were both conjectured by Colin de Verdi{\`e}re in \cite{Colin},
the proofs are similar to the one we gave for the isochore Morse lemma.
 \begin{theorem}[\cite{isochore}] A deformation
$F:(\CM^k \times \CM^2,0) \to (\CM,0)$ of a holomorphic function germ
$f:(\CM^2,0) \to (\CM,0)$ is isochore versal if and only if the Brieskorn lattice of $f$ is generated by
the classes of the forms $(\d_{\l_i}F) dx \w dy$ restricted to $\l=0$.
\end{theorem}
\begin{theorem}[\cite{isochore,quantique}] Under the assumptions of the previous theorem,
the map $\psi:(\CM^j\times \CM,0) \to (\CM^k \times  \CM,0)$
 inducing the deformation $G$ from the isochore versal deformation 
 $F$ in the commutative diagram \ref{E::commutative}
 is uniquely determined by the choices of $F$ and $G$ provided that $F$ is miniversal. 
 \end{theorem}

%%%%%%%%%%%%%%%%%%%%%%%%%%%%%%%%%%%%%%%%%%%%%%%%{\`u}{\`u}
\section{The local finiteness theorem}
The origin of finiteness theorems goes back to Cartan-Serre and the Riesz-Schwartz theory for
compact operators \cite{Cartan_Serre,Schwartz}.
There is now an extensive literature on the subject \cite{Douady,Forster_Knorr,Houzel,Verdier,Schneiders}.
We give a practical criterion -close in spirit to  \cite{Buchweitz,Houzel_Schapira,vanStraten}-
in view of applications to singularity theory.\\
We shall use the notions introduced by Thom and Whitney of stratified spaces and mappings, we refer to \cite{Gibson_Looijenga} and reference therein for details.\\
The notions of standard representative admit a straightforward generalisation for map-germs satisfying Thom's $a_f$ condition. As the stratification that we shall use in practise are rather simple, we will not recall these definitions.
A detailed exposition can be found in \cite{finitude}.
%%%%%%%%%%%%%%%%%%%%%%%%%%%%%%%%%%%%%%%%%%%%%%%%%%%%%%%%%%%%%%%
\subsection{Motivation: the quantum Morse lemma}
As we saw previously, the existence of a versal deformation follows from the finiteness of the associated deformation module.
For the quantum Morse lemma, the problem is to solve the equation
 $$a_t(f_t)+\frac{1}{\hbar} [H_t,f_t] =g.$$
where $a_t \in \Cq \{t,z\}$ and $H_t \in \Qt \{ t \}$ are the unknown.\\
A proof of this almost straightforward statement can be found in \cite{quantique}.
Therefore like for the Vey theorem, we have to show that the module
 $M=\Qt \{ t \}/\frac{1}{\hbar} [\Qt\{ t \},f_t]$ is generated by the class of $1$. 
By making the substitution $t=0$ and taking the semi-classical limit, we get that
$$M/(tM+\hbar M) \approx \Ot_{\CM^2,0}/\{ \Ot_{\CM^2,0} ,f \}$$
which is as shown previously isomorphic to the Brieskorn lattice of $f$. This module is generated by the class of $1$ therefore like for the isochore Morse lemma,
the Nakayama lemma implies that the class of $1$ generates $M$ provided that it is a finite type $\Cq \{ t,z \}$-module.\\ 
Thus, the theorem will follow if we know that the module $M$ is of finite type, that is, from a quantum version of the Brieskorn-Deligne theorem. If moreover the module is free, then the
function $a_t$ is uniquely defined.
%%%%%%%%%%%%%%%%%%%%%%%%%%%%%%%%%%%%%%%%%%%%%%%%%%%%%%%%%%%%5
%%%%%%%%%%%%%%%%%%%%%%%%%%%%%%%%%%%%%%%%%%%%%%%%%%%%%%%%%%%%%%%%%%%5
%%%%%%%%%%%%%%%%%%%%%%%%%%%%%%%%%%%%%%%%%%%%%%%%%5
\subsection{The sheaves $\Ot_{X|Y}$}
Let $i:X \to Y$ be the inclusion of a complex analytic manifold $X$ into another complex analytic manifold $Y$.
The sheaf $i^{-1}\Ot_{Y}$ is denoted by $\Ot_{X|Y}$. If $Y$ is of the form $X \times T$, we denote simply by $\Ot_{X|X \times T}$
the sheaf obtained from the inclusion of $X \times \{ 0 \}$ in $X \times T$. These sheaves are frequently considered in
microlocal analysis (\cite{SKK}).\\
The stalk of  the sheaf $\Ot_{X|X \times T}$ at a point $x_0$ is the space of germs of holomorphic functions in $X \times T$
at the point $x=x_0,\ t=0$.\\
It follows from Cartan's theorem A that the ring  $\Ot_{X|X \times T}$ is coherent, that is, the kernel of any map
$$\Ot^k_{X|X \times T} \to \Ot_{X|X \times T}$$
is finitely generated. Following Serre \cite{Serre_FAC}, we say that a sheaf $\Ft $ on a space $X$ is {\em $\Ot_{X|X \times T}$-coherent}
or that it is a {\em coherent ind-analytic sheaf}, if it is the cokernel of a morphism of $\Ot_{X|X \times T}$-modules:
$$\Ot_{X|X \times T}^{p} \to \Ot_{X|X \times T}^{n} \to \Ft \to 0.$$ 
  %%%%%%%%%%%%%%%%%%%%%%%%%%%%%%%%%%%%%%%%%%%%%%%%%%%%{\`u}
\subsection{The finiteness theorem}
The algebra $\Qt$ is the stalk at the origin of a sheaf in $\CM^2$, this sheaf is defined by the pre-sheaf
$$U \mapsto \Qt_{\CM^{2}}(U)=\{ f \in \Qt:Bf \in \Ot_{\CM^2|\CM^3}(U) \}$$
where $U \subset \CM^{2}$ is an open subset.\\
As a sheaf of vector spaces $\Qt_{\CM^2}$ is isomorphic to the sheaf $ \Ot_{\CM^2|\CM^3}$ but the algebra structure are different
since $B(fg) \neq BfBg$, in general.\\
Let us consider now the case with parameters.\\
The {\em quantum analytic sheaf relative to the projection $(\l,x,y) \mapsto
  \l$}, denoted $\Qt_{\CM^{k+2}/\CM^k}$, is defined by the presheaf  (\cite{Pham_resurgence,Schapira_Polesello}):
$$ U \to \Qt_{\CM^{k+2}/\CM^k}(U)=\{ f \in \widehat \Qt[[\l]] , Bf \in
 \Ot_{\CM^{k+2}|\CM^{k+3}}(U) \}$$
where $U$ denotes an open subset.
The sheaf of vector spaces $\Qt_{\CM^{k+2}/\CM^k}$ and $\Ot_{\CM^{k+2}|\CM^{k+3}}$
are isomorphic.\\ 
The sheaf $\Bt_S$ is defined by  the presheaf:
$$ U \to \Bt_{S}(U)=\{ f \in \hat \CM[[\hbar,\l]] , Bf \in
 \Ot_{\CM^{k}|\CM^{k+1}}(U) \}.$$ 
\begin{definition} Consider a map $F:X \to S$ satisfying Thom's $a_F$ condition.
A  sheaf $\Ft$ is called {\em  $F$-constructible} if the following condition holds:
 for each point $x \in X$ there exists a neighbourhood $U$ inside the strata of $x$
such that
$$\Ft_{|U} \approx F^{-1}(F_{|U})_* \Ft. $$
\end{definition}
A complex of $\Qt_{\CM^{k+2}/\CM^k}$-coherent sheaves is called $F$-constructible it its cohomology sheaves are
$F$-constructible and if its differential is $F^{-1} \Bt_S$ linear. Similar consideration apply for function germs (\cite{finitude}).
\begin{theorem}[\cite{finitude,quantique}]
\label{T::finitude}
 {Let $F:(\CM^k \times \CM^2,0) \to (\CM^l,0)$ be a holomorphic map germ satisfying the $a_F$-condition.
The cohomology spaces $H^k(K^\cdot) $ associated to a complex of $F$-constructible
$\Qt_{\CM^{k+2}/\CM^k,0}$-coherent modules are $F^{-1}\Bt_{\CM^l,0}$-coherent modules.}
\end{theorem}
%%%%%%%%%%%%%%%%%%%%%%%%%%%%%%%%%%%%%%%%%%%%%%%%%%%%%%%5
\subsection{The versality and unicity theorems}
The quantum Morse lemma and the quantum versal deformation theorem are corollaries of the following theorem.
\begin{theorem}[\cite{quantique}]
\label{T::qBrieskorn}The quantised Brieskorn lattice $\Qt\{ \l \}/\frac{1}{\hbar}[f,\Qt]$ associated to an operator $f \in \Qt\{ \l \}$ which symbol is the deformation of an isolated
plane curve singularity is a  finite type $\Cq\{ \l,z \}$-module.
\end{theorem}
To prove the theorem, we consider the unfolding of the principal symbol of $f$:
$$F: (\CM^k \times \CM^2,0) \to (\CM^{k+1},0),\ (\l,x,y) \mapsto (\l,\s(f(\l,x,y)))  $$
and we apply Theorem \ref{T::finitude} to the mapping $F$ and to the complex
$$K^\cdot:0 \to \Qt_{\CM^{k+2}/\CM^k,0} \to \Qt_{\CM^{k+2}/\CM^k,0} \to 0.$$
One easily sees that this complex is $F$-constructible (\cite{quantique}).\\
The unicity of perturbative expansions, also conjectured by Colin de Verdi\`ere (\cite{Colin}), 
is a consequence of the following theorem.
 \begin{theorem}[\cite{quantique}]
Under the assumptions of the theorem, the quantised Brieskorn lattice
$\Qt\{ \l \}/\frac{1}{\hbar}[f,\Qt\{ \l \}]$ a is a  free $\Cq\{ \l,z \}$-module.  
\end{theorem}
\begin{theorem}[\cite{quantique}]
Let $F \in \Qt \{ \l \}$ (resp. in $\hat \Qt [[ \l ]]$)
be a miniversal deformation of an
operator $f \in \Qt$.
Let $G$ be a deformation of $f$, so that $G$ is induced
from $F$, that is, $\psi \circ G=\p(F)$, then the function germs $\p(\l_k) \in \Cq\{ \mu \}$ and $\psi \in \Cq \{ \mu,z \}$
are uniquely determined by the choices of $F$ and $G$.
\end{theorem} 
This concludes our study of deformation theory in the $D=1$ Heisenberg algebra, we now turn to the higher dimensional case.
%%%%%%%%%%%%%%%%%%%%%%%%%%%%%%%%%%%%%%%%%%%%%%%%%%%%%%%%%%%%%%%%%%{\`u}
\section{Deformations of singular Lagrangian varieties}
The characteristic variety of a $D=1$ pseudo-differential operator is a plane curve. In the previous sections, we investigated the relation
of the deformation theory of these plane curves and the perturbative expansions associated to the corresponding eigenvalue problem.
In higher dimensions, the characteristic variety is an
involutive variety and for integrable systems, it is in fact a Lagrangian variety. In this chapter, we will study local deformation
theory for Lagrangian singularities, for simplicity we consider only symplectic manifolds, the generalisation to Poisson manifolds is
straightforward.
%%%%%%%%%%%%%%%%%%%%%%%%%%%%%%%%%%%%%%%%%%%%%%%%%%%%%%%%%%%%%{\`u}
%%%%%%%%%%%%%%%%%%%%%%%%%%%%%%%%%%%%%%%%%%%%%%%%%%%%%%%%%%%%%%%%%%%{\`u}{\`u}
\subsection{The Lagrangian complex}
Deformation theory for singular Lagrangian varieties was settled by Sevenheck and van Straten in \cite{VS}
(see also \cite{Pham_Legendre,Sev_these,Stolovitch}).
In this paper the authors introduce a complex similar to the Chevalley complex of a Lie algebra. We review this construction for the case of integrable systems.\\ 
We endow $\CM^{2n}$ with the holomorphic symplectic two-form $\sum_{i=1}^{2n}dq_i \w dp_i$; by Darboux theorem locally any symplectic manifold is of this type (\cite{MMCM}).\\
A Lagrangian variety in $\CM^{2n}$ is a reduced analytic space of pure dimension $n$ for which the symplectic form vanishes on the smooth locus.\\
A map $f=(f_1,\dots,f_n):(\CM^{2n},0) \to (\CM^n,0)$ is called a {\em Lagrangian singularity} if the ideal $I$ generated by its component defines
the germ of a complete intersection Lagrangian variety germ\footnote{The words {\em Lagrangian singularity}
are also used in the theory of caustics and wave fronts for a different object.};
this condition is equivalent to stating that the Poisson bracket induces a map
$$I \times I \to I,\ (f,g) \mapsto \{ f,g \}.$$ 

Let $f:(\CM^{2n},0) \to (\CM^n,0)$ be a moment mapping and denote by $(L,0)$ the fibre of $f$ above the origin.
As $L$ is a complete intersection, the complex $(C_{L,0}^\cdot,\dt)$ which terms are
$C^k_{L,0}=\Hom(\bigwedge^k \Ot^n_{L,0},\Ot_{L,0})$, might be identified with $\bigwedge^k \Ot^n_{L,0}$.\\
The differential of the complex is defined by the conditions
\begin{enumerate}
\item it is $\CM$-linear
\item for any $v \in \bigwedge \CM^n$ and any $m \in \Ot_{L,0}$, we have
$\dt(mv)=\sum \{ m,e_i \}\w v$ where $e_1,\dots,e_n$ denotes the canonical basis in $\CM^n$.
\end{enumerate}
If $(L,0)$ is smooth then the complex is a resolution of the constant sheaf $\CM_L$.\\
In arbitrary dimension, the cohomology space $H^1(C_{L,0}^\cdot)$ can be canonically identified with the space of infinitesimal Lagrangian deformation modulo symplectic change of coordinates. A proof of these elementary facts is given in \cite{VS}.\\
For $n=1$, there is only one differential given by
$$\Ot_{L,0} \to \Ot_{L,0},\ H \mapsto \{ H,f \} $$
and the vector space $H^1(C_{L,0}^\cdot)$ is canonically isomorphic to the fibre at the origin of the Brieskorn lattice
$$H^1(C_{L,0}^\cdot) \approx H''_f/fH''_f.$$
Consider now the case $n=2$.\\
Let $(F_1,F_2)=(f_1+tm_1,f_2+tm_2)$ be a deformation of the ideal of $(L,0)$ above $Spec(\CM[t]/t^2)$.\\
The condition $\{ F_1, F_2 \}=0$ is equivalent to $\{ m_1,f_2\}+\{f_1,m_2\}=0$, that is, $\dt(m)=0$ with $m=(m_1,m_2)$.\\
Assume that $F$ is a symplectically trivial deformation: this means that there exists a family of symplectomorphisms
$\p_t=Id+t \{H,\cdot \}$ such that $f \circ \p_t= F$, that is, $f+t \{H,f \}=f+tm$ therefore
$m=\dt(H)$. This proves the assertion.\\
The Sevenheck-van Straten construction naturally extends to the relative case, i.e., to a deformation of a Lagrangian variety rather than for a single Lagrangian variety.
Let us now consider this case.\\
 A diagram
 $$\xymatrix{X \ar^-{i}[r] & \CM^k \times \CM^{2n} \ar^{\pi}[r]& S}$$
where $i$ is the inclusion and $\pi$ the projection is called a {\em Lagrangian deformation} if the fibres of
$\pi_{\mid X}$ project to Lagrangian varieties in $\CM^{2n}$.\\
The germ of such a diagram at the origin is defined by a holomorphic map-germ
$F:(\CM^k \times \CM^{2n},0) \to (\CM^n,0)$ which components
generate the ideal of the germ $(X,0)$ in $(\CM^k \times \CM^{2n},0)$.\\
We will use the notations $F:(\CM^k \times \CM^{2n},0) \to (\CM^n,0)$ for germs of Lagrangian deformation
and (abusively) $F:X \to S$ for standard representatives of them,
the corresponding Lagrange complex will be denoted by $C^\cdot_{F}$.
 
%%%%%%%%%%%%%%%%%%%%%%%%%%%%%%%{\`u}
\subsection{Finiteness theorem}
Let $f:(\CM^{2n},0) \to (\CM^n,0)$ be a Lagrangian singularity and consider the variety germs
$(S_k,0)=(\{ x \in L, {\rm rank}\, d f(x)=k \},0)$ where $L$ is the zero fibre of $f$.
\begin{definition}The Lagrangian singularity $f:(\CM^{2n},0) \to (\CM^n,0)$ is called
{\em pyramidal} provided that $\dim S_k \leq k$.
\end{definition}
Vey proved that the Lagrangian singularity $f=(f_1,\dots,f_n):(\CM^{2n},0) \to (\CM^n,0)$ defined by
$f_i=p_iq_i+r_i,\ r_i \in \Mt^3$ is pyramidal\footnote{Although he never considered the notion explicitly.} (\cite{Vey}). \\
The Henon-Heiles integrable system
$$H_1=p_1^2+p_2^2-4q_2^3-2q_1^2q_2,\ H_2=q_1^4+4q_1^2q_2^2-4p_1(q_1p_2-q_2p_1)  $$
defines a pyramidal map.\\
I do not know any example of a map which defines a non-pyramidal Lagrangian complete
intersection. A Lagrangian deformation will be called {\em pyramidal} if it is the deformation of a pyramidal Lagrangian variety.\\
The local Grauert theorem implies the following result.
\begin{theorem}[\cite{lagrange}]
\label{T::Lfinite}
The cohomology spaces $H^k(C^\cdot_F)$ associated to a Lagrangian deformation $F:(\CM^k \times \CM^{2n},0) \to
(\CM^n,0)$  of a pyramidal Lagrangian singularity are finite type $\Ot_{\CM^k,0}$-modules.
\end{theorem}
For the absolute case ($k=0$), the theorem was proved by Sevenheck and van Straten (\cite{VS}).

%%%%%%%%%%%%%%%%%%%%%%%%%%%%%%%%%%%55
\subsection{Rigidity theorem}
Consider a Lagrangian deformation  $F:(\CM^k \times \CM^{2n},0) \to (\CM^n,0)$ and denote by
$F_D:(\CM \times \CM^{2n},0) \to (\CM^n,0)$ the restriction of $F$ to a one dimensional vector space $D \subset \CM^k$ (or equivalently
the deformation induced by a linear base change $\CM \to \CM^k$).\\
There is a  Lagrangian Kodaira-Spencer mapping 
$$\theta_D:\CM^k \to H^1(C^\cdot_{F_D}),\ \d_{\l_i} \mapsto [\d_{\l_i}F_{|\l \in D}],\ i=1,\dots,k$$
where we identified the tangent space to $\CM^k$ at the origin with $\CM^k$ itself and also an absolute
version of it
$$\theta_L:\CM^k \to H^1(C^\cdot_{L,0}),\ \d_{\l_i} \mapsto [\d_{\l_i}F_{|\l=0}],\ i=1,\dots,k$$
In view of applications to integrable systems, let us define a Lagrangian deformation  $ F:(\CM^k \times \CM^{2n},0) \to (\CM^n,0) $
to be {\em rigid} if any deformation $ G:(\CM^{k+1} \times \CM^{2n},0) \to (\CM^n,0) $ whose restriction above $\CM^k$ is equal to $F$
is trivial, i.e., it is obtained from $F$ by symplectic change of coordinates depending on the parameters.\\
The correspondence between rigidity and coherence of the deformation module generalises to higher dimensions without any difficulty.
Thus, from theorem \ref{T::Lfinite}, we deduce the following result.
\begin{theorem}[\cite{lagrange}]
\label{T::rigidite}
{Let $ F:(\CM^k \times \CM^{2n},0) \to (\CM^n,0) $
be a pyramidal Lagrangian deformation  of a  Lagrangian complete 
intersection-germ
$(L,0) \subset (\CM^{2n},0)$. Assume that either the absolute Kodaira-Spencer mapping is surjective of 
there exists a one dimensional subspace $D \subset \CM^k$ such that
the Kodaira-Spencer map
$$\bar \theta_D:T_0\CM^k \to H^1(C^\cdot_{F_D})$$
 is surjective.
Then, the Lagrangian deformation $ F $ is rigid.}
\end{theorem}
\begin{example}
Let $F=(F_1,\dots,F_n)$ be defined by $(p_1q_1+\l_1,\dots,p_nq_n+\l_n)$.
Denote by $e_1=(1,0,\dots,0)$, $\dots,\ e_n=(0,\dots,0,1)$ the canonical base in $\CM^n$.
The absolute Kodaira-Spencer map sends the vector $\d_{\l_i}$ to the class of the vector $e_i$.
An elementary computation shows that this map is surjective. This computation was already done in \cite{Pham_Legendre}
from which the authors deduced the rigidity without proof.
\end{example}
%%%%%%%%%%%%%%%%%%%%%%%%%%%%%%%%%%%%%%%%%%%%%%%%%55
\subsection{Freeness and rigidity}
The direct computation of the Lagrangian deformation module might be difficult, therefore we intend to reduce
it to elementary topological considerations.
The following result and its corollary give a practical way for proving the surjectivity of the  Lagrangian Kodaira-Spencer mapping.
 \begin{proposition}[\cite{mutau}]
\label{P::free}For any  pyramidal Lagrangian deformation over a one dimensional base
$F:(\CM \times \CM^{2n},0) \to (\CM^n,0)$ the cohomology space $H^1(C^\cdot_F)$ is a free $\Ot_{\CM,0}$-module.
\end{proposition}
\begin{proof}
The multiplication by the parameter $t$ of the deformation induces an exact sequence of complexes
$$\xymatrix{0 \ar[r]& C^\cdot_F \ar^t[r] & C^\cdot_F \ar[r]& C^\cdot_{L,0} \ar[r] &0}  $$
which induces in turn an exact sequence in cohomology
$$\xymatrix{\dots \ar[r]& H^p(C^\cdot_F) \ar^t[r] &H^p( C^\cdot_F) \ar[r] &H^p( C^\cdot_{L,0}) \ar[r] &\dots}  $$
as the fibres are reduced Lagrangian varieties, the sequence at the $H^0$-level gives
$$\xymatrix{0 \ar[r]& \Ot_{\CM,0} \ar^t[r] &\Ot_{\CM,0} \ar[r]&\CM \ar[r] &\dots}  $$
Therefore, the exact sequence splits at $H^0(C_{L,0}^\cdot)$. This shows that the multiplication by $t$ in $ H^1(C^\cdot_F)$ is injective, thus the module
is free. 
\end{proof}
\begin{corollary}[\cite{mutau}]
\label{C::mutau} For any pyramidal Lagrangian deformation over a one dimensional base, a smooth fibre of a standard
representative has a first Betti number equal to the rank of the module $H^1(C^\cdot_F)$.
\end{corollary}
 \begin{theorem}[\cite{mutau}]
\label{T::free}
Let $F:(\CM^k \times \CM^{2n},0) \to (\CM^n,0)$ be a pyramidal Lagrangian deformation for which
the  Lagrangian Kodaira-Spencer mapping $\theta_D$ is surjective, for some one dimensional vector space $D \subset \CM^k$,
then the cohomology spaces $H^1(C^\cdot_F)$ associated to this Lagrangian deformation are free $\Ot_{\CM^k,0}$-modules.
\end{theorem}
Thus, like for the one dimensional case, if the Lagrangian Kodaira-Spencer mapping is an isomorphism then
the perturbative expansions for the spectrum are uniquely defined, at least in the semi-classical limit. We will come back to this later,  for the moment, we concentrate ourselves in proving or dis-proving the surjectivity of the Lagrangian Kodaira-Spencer mapping.
\begin{example}
\label{E::lagrange} Let us consider the Lagrangian deformation defined by the mapping germ
$$(\CM^2 \times \CM^4,0) \to (\CM^2,0),\ (\l,q,p) \mapsto (q_1^2+p_1^2+\l_1,q_2^2+p_2^2+\l_2) $$
and take $D=\{ (\l_1,\l_2) \in \CM^2, \l_2=\l_1+1 \}.$ The smooth fibre of a standard representative retracts on a
real 2-torus $S^1 \times S^1$, the first Betti number of which is equal to 2.
Corollary \ref{C::mutau} implies that the module $H^1(C^\cdot_{F_D})$
is isomorphic to $\Ot_{\CM,0}^2$. The cohomology classes
$[(1,0)]=\theta_D(\d_{\l_1})$ and $[(0,1)]=\theta_D(\d_{\l_2})$ are obviously independent and not divisible by $t$; therefore
they generate the module $H^1(C^\cdot_{F_D})$. Thus Theorem \ref{T::free} applies,
consequently $H^1(C^\cdot_{F})$ is isomorphic to $\Ot_{\CM^2,0}^2$ with generators the classes of the constant mappings $(1,0)$ and $(0,1)$. This can be proved by a direct computation but it is important, in view of more complicated examples,
that we were able to prove the result without any computation. 
\end{example}
%%%%%%%%%%%%%%%%%%%%%%%%%%%%%%%%%%%%%%%%%%%%%%%%%%%%%%%%
%%%%%%%%%%%%%%%%%%%%%%%%%%%%%%%%%%%%%%%%%%%%%%%%%%%%%%%%%%%%%%%%%

%%%%%%%%%%%%%%%%%%%%%%%%%%%%%%%%%%%%%%%%%%%%%%%%%%%%%%%%%%%%%%%%%%%%{\`u}{\`u}
\section{The Lagrangian Milnor fibre}
In the previous section, we saw that, if true, the surjectivity of the Kodaira-Spencer mapping can be established once we know the
first Betti number of the Milnor fibre.
%%%%%%%%%%%%%%%%%%%%%%%%%%%%%%%%%%%%%%%%%%%%%%%%%%%%%%%%%%%%%%%%%%%%%%%%%%%5
%%%%%%%%%%%%%%%%%%%%%%%%%%%%%%%%%%%%%%%%%%%%%%%%%%%%%%%%%%%%%%%%%%%%%%%%%%%%%%%%%%%%%55
\subsection{The free basis theorem}
The construction of Lefschetz vanishing cycles for isolated complete intersection singularities by Morsification is classical
and was, according to D.T. L\^e, already known to Tyurina as a consequence of the versal deformation theorem (\cite{AVG}).\\
This construction can be adapted for a complete intersection with non-isolated singularities provided that the generic singular fibre has a Morse transverse singular locus (\cite{Le_Oslo}).\\
As a typical example consider the mapping
$$f:\CM^3 \to \CM, (x,y,z) \to x^2+y^2.$$
Above a point $\e \in \RM_{>0}$, the fibre retracts on the real circle $\{ (x,y) \in \RM^2:x^2+y^2=\e \}$. There are only two generators
of the first homology group corresponding to this circle with the two different orientations, so the ambiguity lies only
on the orientation.\\
The zero fibre of $f$ is said to have an $A_1$-transverse singular locus, if at each critical point, the restriction of $f$ to a generic plane has an isolated critical point with a non-degenerate Hessian.\\
Given a holomorphic mapping $f:X \to S,\ S \subset \CM^k$, we denote by $M(f) \subset S$
the set of values for which the fibre has a connected transverse $A_1$ singular locus.\\
We say that $f$ satisfies condition $(M)$ if the set $M(f)$ is an open variety of codimension one in $S$ and dense
inside the set of critical values.
Under this assumption, the set of critical values is an analytic variety called the {\em discriminant}.\\
If in addition $f:X \to S$ is a standard representative of a holomorphic map-germ, we define the {\em Lefschetz vanishing spheres}
by choosing  a fixed generic complex line $D \subset \CM^k$ and by taking
the cycles which vanish at the intersection of $D$ with the discriminant. Up to isotopy
and orientation, these spheres are uniquely defined.
\begin{example} As a trivial -but already interesting- example, let us consider the case
$$f:(\CM^4,0) \to (\CM^2,0),\ (q,p) \mapsto (q_1^2+p_1^2,q_2^2+p_2^2) .$$
The fibre has the homotopy type of a real torus $S^1 \times S^1$, the discriminant consists of the two coordinate lines in $\CM^2$, it has multiplicity two. The line $D=\{ \l_2=\l_1+1 \}$ intersects the discriminant of $f$ at the points $(1,0)$ and $(0,1)$
at each point there is a one-dimensional vanishing cycle.
\end{example}
\begin{proposition}[\cite{monodromy}] Let $f:X \to S$ be a standard representative of a pyramidal Lagrangian deformation
satisfying condition (M) then the Lefschetz vanishing spheres may be oriented so to define isotopy
classes of spheres independent on the choice of the critical point.
\end{proposition} 
The homology class of a {\em Lefschetz vanishing sphere} will be called a {\em Lefschetz vanishing cycle}.
\begin{theorem}[\cite{monodromy}] Let $f:X \to S$ be a standard representative of a pyramidal
Lagrangian deformation germ. Assume that $X$ is smooth and that $f$ satisfies condition
(M) then the Lefschetz vanishing cycles freely generate the first homology group $H_1(V,\ZM)$ of a Milnor fibre of $f$. In particular,
the first Betti number of $f$ equals the multiplicity of the discriminant of $f$. 
\end{theorem}

%%%%%%%%%%%%%%%%%%%%%%%%%%%%%%%%%%%%%%%%%%%%%%%%%%%%%%%%%%%%%%%%%%%%%%%%%%%
\subsection{The Henon-Heiles integrable system}
\label{SS::Henon}
The Henon-Heiles integrable system
$$H_1=\frac{1}{2}(p_1^2+p_2^2)-2q_2^3-q_1^2q_2,\ H_2=q_1^4+4q_1^2q_2^2+4p_1(p_1q_2-p_2q_1)  $$
defines a pyramidal map-germ $H=(H_1,H_2):(\CM^4,0) \to (\CM^2,0)$ satisfying condition (M).\\

The discriminant of a standard representative is the curve: 
$$\S=\{ (\l_1,\l_2) \in S: \l_2(\l_2^3-3^3\l_1^4)=0 \} .$$
This can be computed directly or using the Lax representations, namely the determinant of the Lax matrix
$$L=\begin{pmatrix} -xp_2+p_1q_1 & x^2+2xq_2-q_1^2 \\
-x^3/2+q_2x^2-x(2q_2^2+q_1^2/2)+p_1^2 &xp_2-p_1q_1 \end{pmatrix}$$
 is equal to
$$\det L=\frac{1}{2}x^5-2H_1 x^2-\frac{1}{2}H_2 x. $$
A level set of $H$ is singular precisely when the degree 5 polynomial $\det L$ has a double root (\cite{Gerardy}).\\
The map $f$ defines a two parameter Lagrangian deformation germ of its zero fibre namely
$$F:(\CM^2 \times \CM^4,0) \mapsto (H_1(q,p)-\l_1,H_2(q,p)-\l_2). $$
The discriminant $\S$ has multiplicity $4$ at the origin, therefore the module $H^1(C^\cdot_{F_D})$ is free of rank $4$.
In particular, the associated Lagrangian Kodaira-Spencer is not surjective since the image is the rank two submodule generated by
the classes of $(1,0)$ and $(0,1)$. Thus, the deformation $F$ is not rigid.\\
By way of contrast, the Lagrangian Kodaira-Spencer maps associated to the 4-parameter Lagrangian deformation
$$G:(\t \l,q,p) \mapsto F(\t \l,q,p)+\l_3 (q_2,-2q_1^2)+\l_4(q_1^2+4q_2^2 ,-8q_1^2q_2) $$
are surjective, $\t \l=(\l_1,\l_2,\l_3,\l_4)$. Indeed, the classes of $(1,0)$, $(0,1)$, $(q_2,-2q_1^2)$ and
$(q_1^2+4q_2^2,-8q_1^2q_2)$ are independent and not divisible by $t$, therefore they generate the module
$H^1(C^\cdot_{F_D})$. Theorem \ref{T::free} implies that the classes of the same map-germs in $H^1(C^\cdot_F)$
freely generate this module.
This fact would have been much more difficult to establish by a direct computation.\\
Let us now make some comments on this example.\\
The discriminant of $G$ coincides
with that of the boundary singularity  $B_4$ in Arnold's classification (\cite{Arnold_BCF}).
A variant of the Picard-Lefschetz formula then shows that the monodromy of the vanishing cycles is that of the
$B_4$ Coxeter group (\cite{monodromy}).\\
This fact can also be proved directly using a Painlev\'e-Kovalevskaia type analysis, in order to show that
the smooth fibres are open parts of Prym varieties of the two-fold
covering $(x,y) \mapsto (x^2,y)$ of elliptic curves $y^2+x^4+a_1 x^3+a_2 x^2+a_3 x+a_4$ (\cite{Lesfari}).\\
Finally remark that by Theorem \ref{T::rigidite}, there is always a holomorphic involution on the fibres of a deformation of $G$
extending the involution $(q_1,p_1,q_2,p_2) \mapsto (-q_1,-p_1,q_2,p_2)$.\\
This shows that symmetries may be rigid in the symplectic world of Lagrangian varieties.
%%%%%%%%%%%%%%%%%%%%%%%%%%%%%%%%%%%%%%%%%%%%%%%%%%%%%%%%%%%%%%%%%%%%%%%%%%%%%%%%%%%%%%%%%%%%%%%%%%%%%%%%%%%%%%%
\section{The R\"ussmann-Vey theorem and its generalisations}
\subsection{The classical R{\"u}ssman-Vey theorem}
Denote by $\Mt$ the maximal ideal of the local ring $\Ot_{\CM^{2n},0}$.
Let $\a_1,\dots,\a_n$ be pairwise rationally independent numbers, then for any formal power series $F=\sum_{i=1}^n \a_i p_iq_i+r$, $r \in \Mt^3$, there exists a formal symplectomorphism $\p$ and a formal power series $\psi \in \CM[[z_1,\dots,z_n]]$
such that $F \circ \p(q,p)=\psi(p_1q_1,\dots,p_nq_n)$ (\cite{Birkhoff}). Even if $H$ is analytic, this series can be divergent
(\cite{Siegel}). The R\"ussman-Vey theorem asserts that it converges provided that there exists a set of independent
Poisson commuting holomorphic functions $F=H_1,\dots,H_n$ whose quadratic part generate a Cartan subalgebra
(maximal and commutative) of the symplectic group.
\begin{theorem}[\cite{Russmann, Vey}]
\label{T::RV}
Let $H=(H_1,\dots,H_n):(\CM^{2n},0) \to (\CM^n,0)$ be a holomorphic map with Poisson-commuting components.
Assume that  $H_i=p_iq_i+r_i$ with $r_i \in \Mt^3$ then there exists biholomorphic map-germs $\p:(\CM^{2n},0) \to
(\CM^{2n},0)$, $\psi:(\CM^n,0) \to (\CM^n,0)$ such that $\p$ is
symplectic and $H \circ \p=\psi(p_1q_1,\dots,p_nq_n)$.
\end{theorem}
For $n=1$, we recover the isochore Morse lemma. This theorem has been
generalised to arbitrary integrable Hamiltonians with Morse critical points at the origin (\cite{Nguyen_Birkhoff}) (after numerous contributions that we do not list here).\\
A map $H=(H_1,\dots,H_n):(\CM^{2n},0) \to
  (\CM^n,0)$ with Poisson commuting components will be called a {\em moment map}. A deformation of a moment mapping having Poisson commuting components
  will be called an {\em integrable deformation}.
A moment map $H=(H_1,\dots,H_n):(\CM^{2n},0) \to  (\CM^n,0)$ will be called {\em $M$-stable} if it admits only
symplectically trivial integrable deformations, that is, for any deformation
$G:(\CM \times \CM^{2n},0) \to (\CM^n,0)$ there exists a commutative diagram
$$\xymatrix{(\CM \times \CM^{2n},0) \ar[r]^{\t G} \ar[d]^\p& (\CM \times \CM^n,0) \ar[d]^\psi\\
                    (\CM^{2n},0) \ar[r]^F& (\CM^n,0)} $$
where $\t G$ denotes the unfolding of the deformation and $\p$ is a holomorphic map which
preserve the symplectic form $\sum_{i=1}^n dq_i \w dp_i$.\\
Vey observed that under the assumptions of Theorem \ref{T::RV}, the path method applies (\cite{Vey}).
Thus, this theorem is a consequence of the following more general statement.
\begin{theorem}
\label{T::RV2}
Any moment mapping 
$H=(H_1,\dots,H_n):(\CM \times \CM^{2n},0) \to  (\CM^n,0)$ with
$H_i=p_iq_i+r_i,\ r_i \in \Mt^3$ is $M$-stable.\end{theorem}

%%%%%%%%%%%%%%%%%%%%%%%%%%%%%%%%%%%%%%%%%%%%%%%%%%%%%%%%%%%%%%%%%%%%%%%%%%%%%%%%%%5
\subsection{The generalised R\"usmann-Vey theorem}

To a moment mapping $H:(\CM^{2n},0) \to (\CM^n,0)$, we associate the Lagrangian deformation
$$\t H:(\CM^n \times \CM^{2n},0) \to (\CM^n,0),\ (\l,q,p) \mapsto H(q,p)-\l.$$
We say that the moment mapping is {\em pyramidal} if its zero fibre is a pyramidal Lagrangian complete intersection.\\
We will denote by $e_1=(1,0,\dots,0),\dots,e_n=(0,\dots,0,1)$ be the canonical basis in $\CM^n$.\\
The following theorem is a generalisation of the R\"usmann-Vey theorem.
\begin{theorem}[\cite{moment}]
\label{T::RV_general} A pyramidal moment mapping germ $H:(\CM^{2n},0) \to (\CM^n,0)$
is $M$-stable if and only if $H^1(C_{\t H})^\cdot$ is generated by the classes of $e_1,\dots,e_n$.
\end{theorem} 
For corank one maps, this correspondence was stated in \cite{Colin} without proof.\\
Recall that a deformation over a one dimensional base is called a {\em smoothing} if all fibres except the special one are smooth.
\begin{theorem}[\cite{moment,mutau}]
\label{T::Mather}
Let $H:(\CM^{2n},0) \to (\CM^n,0)$ be a pyramidal moment mapping germ, denote by $F:(\CM \times \CM^{2n},0) \to (\CM^n,0)$
an arbitrary one-parameter Lagrangian deformation associated to $H$, that is, $F(t,q,p)=H(q,p)+tv$ with $v \in \CM^n$.
Then, the following conditions are equivalent
\begin{enumerate}
\item The $\Ot_{\CM^n,0}$-module  $H^1(C_{\t H}^\cdot)$ is generated by the classes of $e_1,\dots,e_n$,
\item  The $\Ot_{\CM,0}$-module $H^1(C_{F}^\cdot)$ is generated by the classes  of $e_1,\dots,e_n$.
\end{enumerate}
\end{theorem} 
\begin{proof}
If the classes of $e_1,\dots,e_n$ generate the space $H^1(C_{F}^\cdot)$ then $H^1(C_{\t H}^\cdot)$ is the free module equal to
$\Ot_{S,0} \otimes H^1(C_{F}^\cdot)$ (Theorem \ref{T::free}).\\
We now show that $(i) \implies (ii)$.\\ 
The Lagrangian deformation $F$ is of the type $F(t,q,p)=H(q,p)+t\,v $ for some vector $v \in \CM^n$. Chose a linear
hyperplane transversal to the line generated by $v$ and denote by $\Mt$ the submodule generated by the functions
vanishing on this hyperplane.
There is an injection  $H^1(C_{\t H}^\cdot)/\Mt H^1(C_{\t H}^\cdot) \to H^1(C_F^\cdot)$ (\cite{lagrange}, Proposition 1). This implies
that the multiplication by $t$ is injective in $H^1(C_{\t H}^\cdot)/\Mt H^1(C_{\t H}^\cdot)$  and consequently
this module is free.\\
By Nakayama lemma, we have
 $$\rank H^1(C_{\t H}^\cdot) \leq \rank H^1(C_{\t H}^\cdot)/\Mt H^1(C_{\t H}^\cdot) \leq \rank H^1(C_F^\cdot)$$
This injection maps isomorphically the vector space generated by the classes of the $e_i$'s onto its image
$V \subset H^1(C_{L,0}^\cdot) $.\\
For any standard representative $H:X \to S$ of $F$, we have the isomorphisms
$$H^1(C_{H}^\cdot) \approx (\RM^1 H_*C_{H}^\cdot)_0,\ (\RM^1 H_*C_{H}^\cdot)_s \approx H^1(V,\CM)\otimes \Ot_{S,s} $$
at any regular value $s \in S$, where $V$ denotes the Milnor fibre of $f$ (see \cite{mutau}). Since the sheaves $\RM^1 H_*C_{H}^\cdot$
are coherent this shows that the rank of the module $ H^1(C_{\t H}^\cdot)$ is at least equal to $\b_1(V)$. The module
$H^1(C^\cdot_F)$ is free of rank $\b_1(V)$ (Proposition \ref{P::free}), consequently
$$\rank H^1(C_{\t H}^\cdot) \geq \rank H^1(C_F^\cdot).$$
and therefore both modules have the same rank. In particular, there is an isomorphism
$ H^1(C_{\t H}^\cdot)/\Mt H^1(C_{\t H}^\cdot) \approx H^1(C_F^\cdot)$. This proves the theorem.
\end{proof}
\begin{corollary} For any $M$-stable pyramidal moment mapping germ \\ $H:(\CM^{2n},0) \to (\CM^n,0)$,
 the first Betti number of its Milnor fibre is at most equal to $n$.
\end{corollary}
The results of this section show that these theorems imply the classical
R\"usmann-Vey theorem. Indeed, the computation of Example \ref{E::lagrange} shows that Theorem \ref{T::Mather} and consequently Theorem \ref{T::RV_general} applies.\\
Many other examples can be obtained, consider for instance the Lagrangian versal unfolding of the Henon-Heiles constructed in
Subsection \ref{SS::Henon}. This unfolding gives a moment mapping
$F:(\CM^8,0) \to (\CM^4,0)$
defined by
$$ (q,p) \mapsto  H(\t q,\t p)+p_3 (q_2,-2q_1^2)+p_4(q_1^2+4q_2^2,-8q_1^2q_2) $$
 which is $M$-stable. Here $H=(H_1,H_2)$ denote the Henon-Heiles Hamiltonians considered in Subsection \ref{SS::Henon},
$\t q=(q_1,q_2)$ and $\t p=(p_1,p_2)$.\\

It is unknown whether the theorems of this subsection are true in the real $C^\infty$ case or not.
Nevertheless, for the case considered by R\"usmann and Vey, i.e.,
if the complexification of the quadratic part of the Hamiltonian is like in the R\"usmann-Vey theorem,
then both theorems hold (\cite{Miranda}). In the more particular case of an elliptic integrable system, the result is well-known to semi-classical analysts, it was first proved by
Eliasson in \cite{Eliasson}. The difficulty in the $C^\infty$-case being precisely that the absence of a finiteness theory
does give any simple argument for the implication
 $$H^1(K_H^\cdot)=0 \implies H^1(K_G^\cdot)=0 $$
for any deformation $G$ of $H$.

%%%%%%%%%%%%%%%%%%%%%%%%%%%%%%%%%%%%%%%%%%%%%%%%%%%%%%%%%%%%%%%%%%%%%%5
\section{Quantum integrable systems}
The deformation theory for quantum integrable systems was first settled by van Straten in the unpublished work
\cite{vanStraten_quantique}.
In this work a complex similar to the Lagrangian complex was constructed from which the construction of this section is inspired.
%%%%%%%%%%%%%%%%%%%%%%%%%%%%%%%%%%%%%%%%%%%%%%%%%%%%%%%%%%%%%%%%%%%%%%%%%%%%%%%
\subsection{The quantum deformation complex}
The $D=n$ universal Heisenberg algebra is the algebra generated by $2n+1$ elements
$(a_i^\dag), a_i, \hbar$, $i=1,\dots,n$ and the only non-trivial commuting relations are
$$[a_i,a^\dag_i]=\sqrt{-1}\hbar.$$
Like for $D=1$ case, the formal power series having a total symbol with convergent Borel transform form a subalgebra denote $\Qt_n$.
Our notation is slightly abusive since we denoted simply by $\Qt$ the algebra which is now denoted by $\Qt_1$.\\
A {\em quantum integrable system} $H_1,\dots,H_n \in \Qt_n$consists of $n$-commuting elements in $\Qt_n$; an integrable deformation of it is given
by commuting elements $F_1,\dots,F_n \in \Qt_n\{ t \}$ such that $ (F_i)_{\mid t=0}=H_i $.\\
To the integrable quantum deformation $F=(F_1,\dots,F_n)$, we associate a complex $(C_F^\cdot,\dt)$ which is a quantised version of
the Lagrange complex associated to $\s(F)$, where $\s$ denotes the principal symbol.\\
The terms of this complex are given by $C_F^j=\bigwedge^k \Qt \{ t \}$
The differential $\dt:C_F^j \to C_F^{j+1}$ is defined by the conditions
\begin{enumerate}
\item it is $\Cq \{ t ,z_1,\dots,z_n \}$-linear,
\item for $m \in \Qt \{ t \}$ and $v \in \bigwedge^j \CM^n$, $\dt(m,v)=\sum_{k=1}^n\frac{1}{\hbar} [ m, F_k ] v \w e_k$.
\end{enumerate}

%%%%%%%%%%%%%%%%%%% %%%%%%%%%%%%%%
\subsection{Quantum R\"usmann-Vey theorem}
Like for the $D=1$, the following result follows from the finiteness
theorem (Theorem \ref{T::finitude}). 
\begin{theorem} Let $F=(F_1,\dots,F_n)$ be an integrable quantum deformation.
Assume that the principal symbol of $F$ defines a pyramidal Lagrangian singularity.
Then the cohomology spaces $H^k(C^\cdot_F)$ are finite type $\Cq \{ z_1,\dots,z_n,t
\}$-modules.
\end{theorem}
\begin{theorem} Let $F=(F_1,\dots,F_n)$ be an integrable quantum deformation of a
quantum integrable system $H=(H_1,\dots,H_n) $ and assume that
the symbol of $H$ is pyramidal and $M$-stable then $F=(F_1,\dots,F_n)$ is a trivial deformation.
\end{theorem}
\begin{proof}
The relation between deformations and perturbative expansions implies that it is sufficient to prove that $F$ is a trivial
deformation.
Like for the $D=1$ case, the path method leads to a system of equations of the type
 $$a_{i,t} \circ F+\frac{1}{\hbar}[G_t,F_i]=-\frac{d}{dt}F_i,\ i=1,\dots,n$$
where the $a_{i,t}$ and $G_t$ are the unknown.\\
Denote respectively by $V,\ \t V,\bar V$, the submodule generated by the
classes of the constant mappings in $H^1(C_F^\cdot),\ H^1(C_H^\cdot),\ H^1(C_{\s(H)}^\cdot)$
where $\s(H)$ denotes the $n$-parameter Lagrangian deformation associated to the symbol of $H$.\\
The infinitesimal equations can be solved provided that the submodule $V$ generated by the
classes of the constant mappings is equal to $H^1(C_F^\cdot)$.\\
As $H^1(C_F^\cdot)$ is a finite type module, by Nakayama's lemma, we have an equivalence
$$H^1(C_F^\cdot)=V \iff H^1(C_H^\cdot)= \bar V,$$
The multiplication by $\hbar$ gives a short exact sequence
$$\xymatrix{0 \ar[r] & C_H^\cdot \ar^{\hbar}[r] & C_H^\cdot \ar[r] & C_{\s(H)}^\cdot \ar[r] & 0}$$
This short exact sequence induces a long exact sequence in cohomology which splits at the $H^0$-level:
$$ 0 \to H^1(C_H^\cdot) \to H^1(C_H^\cdot) \to H^1(C_{\s(H)}^\cdot) \to \cdots .  $$
As $ H^1(C_H^\cdot)$ is a module of finite type, by the Nakayama lemma, we get the implication
$$   H^1(C_{\s(H)}^\cdot)=\bar V  \implies  H^1(C_H^\cdot)=\t V $$
Finally, the equality $  H^1(C_{\s(H)}^\cdot)=\bar V $ holds provided that $H^1(K^\cdot_{\s(H)})=0$. 
This proves the theorem.
\end{proof}
This theorem implies that the perturbative expansions of the spectrum are Gevrey-convergent for operators which have an $M$-stable symbol
and if the classes of the constants freely generate the space $H^1(C^\cdot_H)$ then these expansions are unique.
%%%%%%%%%%%%%%%%%%%%%%%%%%%%%%%%%%%%%%%%%%%%%

%%%%%%%%%%%%%%%%%%%%%%%%%%%%%%%%
%%%%%%%%%%%%%%%
\bibliographystyle{amsplain}
\bibliography{master}
\end{document}